\pdfoutput=1
\documentclass{raa}
\usepackage{graphicx,times,amsmath}
 \usepackage{ifpdf}

\begin{document}

\title{Attitude dynamics and control of spacecraft using geomagnetic Lorentz
force}
\author{Yehia A. Abdel-Aziz \inst{1},\inst{2} \and Muhammad Shoaib \inst{2} }
\date{Received~~2009 month day; accepted~~2009~~month day}

\volnopage{Vol.0 (200x) No.0, 000--000}
\setcounter{page}{1} 
\institute{ National Research Institute of Astronomy and Geophysics (NRIAG), Helwan, Cairo, Egypt;
{\it yehia@nriag.sci.eg}\\
\and
University of Hail, Department of Mathematics, PO BOX
2440, Kingdom of Saudi Arabia, {\it safridi@gmail.com}}
\abstract{  The attitude stabilization of a charged rigid spacecraft in Low Earth Orbit (LEO) using torques
due to Lorentz force in pitch and roll directions is considered. A
spacecraft that generates an electrostatic charge on its surface in the
Earth magnetic field will be subject to perturbations from Lorentz force.
The Lorentz force acting on an electrostatically charged spacecraft may
provide a useful thrust for controlling a spacecraft's orientation. We
assume that the spacecraft is moving in the Earth's magnetic field in an
elliptical orbit under the effects of the gravitational, geomagnetic and
Lorentz torques. The magnetic field of the Earth is modeled as a non-tilted
dipole.  A model incorporating all Lorentz torques as a function of orbital
elements has been developed on the basis of electric and magnetic fields.
The stability of the spacecraft orientation is investigated both
analytically and numerically. The existence and stability of equilibrium
positions is investigated for different values of the charge to mass ratio
($\alpha^*$). Stable orbits are identified for various values of $\alpha^*$.
The main parameters for stabilization of the spacecraft are $\alpha^*$ and
the difference between the components of the moment of inertia of spacecraft.  }
\authorrunning{Yehia A. Abdel-Aziz \& M. Shoaib }
\titlerunning{Attitude dynamics and control of spacecraft using geomagnetic
Lorentz force }

\maketitle

\section{Introduction}

The attitude stabilization of a spacecraft is subject to the perturbation
torques which produce turning moments about the center of mass of an
orbiting spacecraft. The significant effect of these torque disturbances on
the spacecraft is dependent on the configuration of the spacecraft.The
perturbation torques may be used to produce a persistent turning moment
about the center of mass of the spacecraft.

The present work analyze the attitude stabilization of a charged spacecraft
by taking into account the effects of gravitational torque, geomagnetic
torque and Lorentz torque. In the case of electrostatically charged
spacecraft, due to the interaction with space plasma, the Lorentz force must
be taken into account as a perturbation on the orbital and attitude motions
of the spacecraft.
The nascent concept of Lorentz spacecraft which is an electrostatically
charged space vehicle may provide a new approach into the solution of the
attitude stabilization of a spacecraft moving around the Earth in low Earth
orbit (LEO). Recently a novel attitude orientation and formation flying
concept using electrostatic propulsion has been proposed by Pollock et al.
(2011), and Chad and Yang (2012). The charge of the spacecraft is controlled
to generate inter-spacecraft Coulomb forces in geostationary orbit. Lorentz
force is a possible means for charging and thus controlling the spacecraft
orbits without consuming propellant (Hiroshima et al. 2009). Peck (2005) was
the first to introduce a control scheme using Lorentz augmented orbits. The
spacecraft orbits accelerated by the Lorentz force are termed Lorentz
-augmented orbits, because Lorentz force cannot completely replace the
traditional rocket propulsion. Many authors introduced Lorentz force as
perturbations on the orbital motion and formation flying such as in
Vokrouhlicky (1989), Abdel-Aziz (2007a), Streetman and Peck (2007),
Hiroshima et al. (2009) , Gangestad et al. (2010), and Abdel-Aziz and Khalil
(2014).

Abdel-Aziz (2007b) studied the attitude stabilization of rigid spacecraft
moving in a circular orbit due to Lorentz torque in the case of uniform
magnetic field and cylindrical shape of spacecraft. Yamakawa et al. (2012)
investigated the attitude motion of a charged pendulum spacecraft moving in
circular orbit, having the shape of a dumbbell pendulum due to Lorentz
torque. Their analysis of the stability of the equilibrium points are
focused only on pitch direction within the equatorial plane. In a recent
study Abdel-Aziz and Shoaib (2014) studied the relation between the
magnitude of Lorentz torque and inclination of the orbits for certain
equilibrium positions where the spacecraft was considered to be in circular
orbit.

In this paper, we analyze the attitude stabilization of a charged spacecraft
moving in geomagnetic field in Low Earth Orbit (LEO). We developed a new
model for the torque due to the Lorentz force for the general shape of the
spacecraft using the Earth magnetic field, which is modeled as a non-tilted
diploe. The total Lorentz force and its torque are developed as a function
of orbital elements of the spacecraft. A dynamical model is built to
describe the attitude dynamics of Lorentz spacecraft. Therefore, based on
the dynamical model, the required control torque due to Lorentz force for
different configurations is developed. The Lorentz acceleration can't
compensate the total propellant but can be used to reduce the consumption of
propellant. Thus, this paper analyzes the attitude stability of the
spacecraft with the Lorentz acceleration and gives the corresponding
required specific charge to mass ratio for such attitude orientation. This
paper also analyzes the effects of charge to mass ratio on the position and
stability of equilibrium positions. We also numerically analyze the behavior
of orbits close to the equilibrium positions.

\subsection{Formulation of the Spacecraft}

We assume that the spacecraft is equipped with an electrostatically charged
protective shield, having an intrinsic magnetic moment. The attitude
orientation of the spacecraft about its center of mass is analyzed under the
influence of gravity gradient torque ${T}_{G}$ , Magnetic torque ${T}_{M}$
and the torque ${T}_{L}$ due to Lorentz force. The torque ${T}_{L}$ results
from the interaction of the geomagnetic field with the charged screen of the
electrostatic shield.

We consider the orbital coordinate system $C_{{x_{o}}{y_{o}}{z_{o}}}$ with $%
C_{x_{o}}$ tangent to the orbit in the direction of motion, $C_{y_{o}}$ lies
along the normal to the orbital plane, and $C_{z_{o}}$ lies along the radius
vector $r$ of the point $O_{E}$ relative to the center of the Earth. The
investigation is carried out assuming the rotation of the orbital coordinate
system relative to the inertial system with the angular velocity $\Omega $.
As an inertial coordinate system, the system $O_{XYZ}$ is taken, whose axis $%
OZ(k)$ is directed along the axis of the Earth's rotation, the axis $OX(i)$
is directed toward the ascending node of the orbit, and the plane coincides
with the equatorial plane. Also, we assume that the spacecraft's principal
axes of inertia $C_{{x_{b}}{y_{b}}{z_{b}}}$ are rigidly fixed to a
spacecraft $(i_{b},j_{b},k_{b})$. The spacecraft's attitude may be described
in several ways, in this paper the attitude will be described by the angle
of yaw $\psi \,$ the angle of pitch $\theta $ , and the angle of roll $%
\varphi $, between the spacecraft's $C_{{x_{b}}{y_{b}}{z_{b}}}$ and the set
of reference axes $O_{XYZ}$. The three angles are obtained by rotating
spacecraft axes from an attitude coinciding with the reference axes to
describe attitude in the following way:

- The angle of precession $\psi \,$is taken in plane orthogonal to $Z$-axis.

- $\theta $ is the rotation angle between the axes $Z$ $\ $and ${z_{0}.}$

- ${\phi }$ is angle of self -rotation around the $Z$-axis

We write the relationship between the reference frames $C_{{x_{b}}{y_{b}}{%
z_{b}}}$ and $C_{{x_{o}}{y_{o}}{z_{0}}}$ as below (Wertz, 1978):
\begin{equation}
A=\left(
\begin{array}{ccc}
\alpha _{1} & \alpha _{2} & \alpha _{3} \\
\beta _{1} & \beta _{2} & \beta _{3} \\
\gamma _{1} & \gamma _{2} & \gamma _{3}%
\end{array}%
\right) ,
\end{equation}

where

\begin{equation}
\begin{array}{l}
({\alpha _{1},\alpha }_{2},\alpha _{3})={(\cos \psi \cos \phi -\sin \psi
\sin \phi \cos \theta ,-\,\cos \psi \sin \phi -\cos \theta \sin \psi \cos
\phi ,\sin \theta \,\sin \psi ),} \\
({\beta _{1},\beta _{2},\beta _{3})=(\sin \psi \cos \varphi \,+\,\,\cos
\theta \cos \psi \sin \varphi ,-\sin \psi \sin \phi +\cos \theta \cos \psi
\cos \phi ,-\,\sin \theta \cos \psi ),} \\
{\,(\gamma _{1},\gamma _{2},\gamma _{3})=(\sin \theta \,\sin \phi ,\sin
\theta \cos \phi ,\cos \theta ),}%
\end{array}%
\end{equation}

and
\begin{equation}
\vec{\alpha}=\alpha _{1}i_{b}+\alpha _{2}j_{b}+\alpha _{3}k_{b}{,}\,\,\,{%
\vec{\beta}}=\beta _{1}i_{b}+\beta _{2}j_{b}+\beta _{3}k_{b}{,}\,\,{\vec{%
\gamma}}=\gamma _{1}i_{b}+\gamma _{2}j_{b}+\gamma _{3}k_{b}.\,
\end{equation}

\begin{figure}[t]
\centering
\resizebox{70mm}{!}{\includegraphics {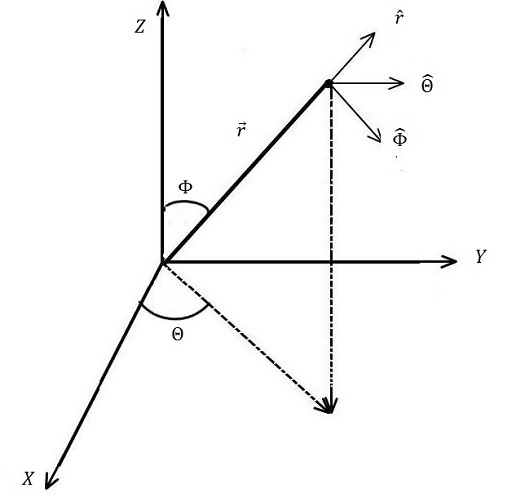}}
\caption{Spherical coordinates used in the derivation of the equations of
motion}
\label{Fig1}
\end{figure}
\section{Total Torque due Lorentz Force}

We use spherical coordinates to describe the magnetic and gravitational
fields, and the spacecraft trajectory, as shown in Figure (1). The $X$, $Y,$
and $Z$ axes form a set of inertial cartesian coordinates. The Earth is
assumed to rotate about the $Z$-axes. The magnetic dipole is not tilted and
therefore, axi-symmetric. The spherical coordinates consist of radius r,
colatitude angle $\Phi $, and azimuth from the $X$ direction $\Theta $ (see
Figure. 1). The magnetic field is expressed as%
\begin{equation}
\vec{B}\mathbf{=}\frac{B_{0}}{r^{3}}\left[ 2\cos \Phi ~\hat{r}\mathbf{+}\sin
\Phi ~\hat{\Phi}\mathbf{+0}\hat{\Theta}\right] ,
\end{equation}%
where $B_{0}$ is the strength of the magnetic field in Wb m. The
acceleration in inertial coordinates is given by

\begin{equation}
\vec{a}\mathbf{=}\frac{\vec{F}}{m}=-\frac{\mu }{r^{3}}\vec{r}\mathbf{+}\frac{%
q}{m}(\vec{E}+\vec{V}_{rel}\times \vec{B}\mathbf{),}
\end{equation}%
where $\frac{q}{m}$is the charge-to-mass ratio of the spacecraft, $\vec{V}%
_{rel}$ is the velocity of the spacecraft relative to the magnetic field of
The Earth. The total Lorentz force (per unit mass) can be written as:

\begin{eqnarray}
\vec{F_{L}} &=&\frac{q}{m}\left[ \vec{E}+\vec{V}_{rel}\times \vec{B}\right] =%
\frac{q}{m}\vec{E}+\frac{q}{m}(\vec{V}_{rel}\times \vec{B}\mathbf{)}  \notag
\\
&=&\vec{F}_{elec}+\vec{F}_{mag},
\end{eqnarray}%
where $\vec{F}_{mag}$ is the Lorentz force experienced by magnetic field and
$F_{elec}$ is the Lorentz force experienced by an electric dipole moment in
the presence of electric field,

\begin{equation}
\vec{F}_{elec}=\frac{q}{m}\vec{E}.
\end{equation}%
Now we start with $\vec{F}_{mag}$ using Maxwell (1861), we can write

\begin{equation}
\vec{F}_{mag}=\frac{q}{m}(\vec{V}_{rel}\times \vec{B}\mathbf{),}\vec{V}%
_{rel}=\vec{V}-\vec{\omega _{e}}\times \vec{r},
\end{equation}%
where $\vec{V}$ is the inertial velocity of the spacecraft, $\mathbf{\omega }%
_{e}~$is the angular velocity vector of the Earth. According to Gangestad et
al. (2010), we used

\begin{equation*}
\vec{V}\mathbf{=}~\dot{{r}}~\hat{r}\mathbf{+}{r}\dot{\Phi}\hat{\Phi}\mathbf{+%
}{r}\mathbf{~}\dot{\Theta}\sin {\Phi }~\hat{\Theta},
\end{equation*}%
and

\begin{equation}
\vec{r}\mathbf{~=}r\mathbf{~}\hat{r},~\vec{\omega _{e}}=\omega _{e}~\hat{z},~%
\hat{z}~=\cos \Phi ~\hat{r}\mathbf{+}\sin \Phi \mathbf{~}\hat{\Phi}.
\end{equation}%
Therefore the acceleration in inertial coordinates is given by

\begin{equation}
\vec{F}_{mag}=\frac{qB_{0}}{m~r^{2}}\left[
\begin{array}{c}
-\left( \dot{\Theta}-\omega _{e}\right) \left( \sin ^{2}\Phi ~\hat{r}+\sin (2%
{\Phi })\hat{\Phi}\right) \\
+\left( \frac{\dot{r}}{r}\sin \Phi -2\dot{\Phi}~\cos \Phi \right) \hat{\Theta%
}%
\end{array}%
\right] .
\end{equation}%
In the case of Torque we need the perturbing force $\vec{F}_{L}$ decomposed
into radial, transverse, and normal direction. The unit vector $\hat{n}$
normal to the orbit is collinear with the angular momentum unit vector $\hat{%
h}$.

\begin{equation}
\hat{n}=\hat{h}=\left( \vec{r}\times \vec{V}\right) /\sqrt{\mu p}=r^{2}/%
\sqrt{\mu p}(-\dot{\Theta}\sin \Phi \hat{\Phi}+\dot{\Phi}\hat{\Theta}),
\end{equation}%
where $p=a(1-e^{2})$, $\mu $ is the Earth's gravitational parameter, $a$ is
the semi-major axis, $e$ is the eccentricity of the spacecraft orbit, and
the transverse unit vector $\hat{t}$ can be calculated from the right-handed
set, $\hat{t}=\hat{n}\times \hat{r}$. Decomposition of the Lorentz force
experienced by the geomagnetic field into the radial, transverse, and normal
components $\left( R_{mag},T_{mag},~N_{mag}\right) $ respectively yields,

\begin{equation}
R_{mag}=\vec{F}_{mag}\cdot \hat{r}~=\frac{q}{m}\frac{B_{0}}{~r^{2}}\left[
\omega _{e}-\dot{\Theta}\right] \sin ^{2}\Phi ,
\end{equation}%
\begin{equation}
T_{mag}=\vec{F}_{mag}\cdot \hat{t}~=\frac{q}{m}\frac{B_{0}}{~\sqrt{\mu p}}%
\left[
\begin{array}{c}
\frac{\dot{r}}{r}\dot{\Theta}\sin ^{2}\Phi \\
-2\omega _{e}~\dot{\Phi}\cos \Phi \sin \Phi%
\end{array}%
\right] ,
\end{equation}%
\begin{equation}
N_{mag}=\vec{F}_{mag}\cdot \hat{n}~=\frac{q}{m}\frac{B_{0}}{~\sqrt{\mu p}}%
\left[
\begin{array}{c}
2\dot{\Theta}(\omega _{e}-\dot{\Theta})\sin ^{2}\Phi \cos \Phi ~ \\
+\frac{\dot{r}}{r}\dot{\Theta}\sin \Phi -2\dot{\Phi}^{2}~\cos \Phi%
\end{array}%
\right] .
\end{equation}%
The relationship between the spherical coordinates and the orbital elements
is required to derive the components of Lorentz force experienced by
magnetic part as a function of orbital elements.

\begin{equation}
r=p/\left( 1+e~\cos f\right) ,~\dot{r}=e\sqrt{\mu /p}\sin f,
\end{equation}%
\begin{equation}
\cos \Phi =\sin i\sin \left( \omega ^{\ast }+f\right) ,\sin \Phi =\sqrt{%
1-\sin ^{2}i\sin ^{2}\left( \omega ^{\ast }+f\right) },
\end{equation}%
\begin{equation}
\dot{\Phi}=-\sqrt{\mu /p^{3}}\frac{\sin i\cos \left( \omega ^{\ast
}+f\right) }{\sqrt{1-\sin ^{2}i\cos ^{2}\left( \omega ^{\ast }+f\right) }}%
\left( 1+e~\cos f\right) ^{2},
\end{equation}%
\begin{equation}
\dot{\Theta}=\sqrt{\mu /p^{3}}\frac{\cos i~}{1-\sin ^{2}i\sin ^{2}\left(
\omega ^{\ast }+f\right) }\left( 1+e~\cos f\right) ^{2},
\end{equation}%
where $i$, $\omega ^{\ast }$ and $f$ are the inclination of the orbit on the
equator, argument of the perigee, and the true anomaly of the spacecraft
orbit respectively. Therefore, rewriting the components of the magnetic part
of the Lorentz force as a function of orbital elements, we obtain
\begin{equation}
R_{mag}=\frac{q}{m}\frac{B_{0}}{~r^{2}}\left[
\begin{array}{c}
\omega _{e}(1-\sin ^{2}i\sin ^{2}\left( \omega ^{\ast }+f\right) ) \\
-\sqrt{\mu /p^{3}}\cos i~\left( 1+e~\cos f\right) ^{2}%
\end{array}%
\right] ,
\end{equation}

\begin{equation}
T_{mag}=\frac{q}{m}\frac{B_{0}}{~\sqrt{\mu p}}\left[
\begin{array}{c}
\frac{\dot{r}}{r}\sqrt{\mu /p^{3}}\cos i~\left( 1+e~\cos f\right)
^{2}+2\omega _{e}~\sqrt{\mu /p^{3}}\sin ^{2}i\sin \left( \omega ^{\ast
}+f\right) \\
\times \cos \left( \omega ^{\ast }+f\right) \left( 1+e~\cos f\right) ^{2}%
\end{array}%
\right] ,  \label{26}
\end{equation}

\begin{eqnarray}
N_{mag} &=&\frac{q}{m}\frac{B_{0}}{~\sqrt{\mu p}}\times  \label{27} \\
&&\left(
\begin{array}{c}
2\left( \omega _{e}(1-\sin ^{2}i\sin ^{2}\left( \omega ^{\ast }+f\right) )-%
\sqrt{\mu /p^{3}}\cos i~\left( 1+e~\cos f\right) ^{2}\right) \\
\sqrt{\mu /p^{3}}\cos i~\left( 1+e~\cos f\right) ^{2}~+\frac{\dot{r}}{r}%
\sqrt{\mu /p^{3}}\frac{\cos i~}{\sqrt{1-\sin ^{2}i\sin ^{2}\left( \omega
^{\ast }+f\right) }}\times \\
\left( 1+e~\cos f\right) ^{2}-2\frac{\mu }{p^{3}}\frac{\sin ^{3}i\cos
^{2}\left( \omega ^{\ast }+f\right) \sin \left( \omega ^{\ast }+f\right) }{%
1-\sin ^{2}i\cos ^{2}\left( \omega ^{\ast }+f\right) }\left( 1+e~\cos
f\right) ^{4}%
\end{array}%
\right) .  \notag
\end{eqnarray}

Now we develop the Lorentz force experienced by electric field $F_{elec}.$

According to Ulaby (2005) and Heilmann et al. (2012) we can write the
electric force as follows.

\begin{equation}
\vec{F}_{elec}=-\nabla \vec{V}_{elec}=\left( \frac{\partial ~V_{elec}}{%
\partial r}\hat{r}\mathbf{+}\frac{1}{r}\frac{\partial ~V_{elec}}{\partial
\Phi }\hat{\Phi}\mathbf{+}\frac{1}{r\sin \Theta }\frac{\partial ~V_{elec}}{%
\partial \Theta }\hat{\Theta}\right) ,
\end{equation}%
where $\vec{V}_{elec}$ is the electric potential,

\begin{equation}
V_{elec}=\frac{\vec{P}\cdot \hat{r}}{4\pi ~\epsilon _{o}r^{2}}.
\end{equation}
$\vec{P}=q\vec{d}$ is called the electric dipole moment, $\vec{d}$ is the
distance vector from charge $-q$ to charge $+q$, $\epsilon _{o}=8.85\times
10^{-12}$ coul$^{2}/(N-m^{2})\equiv volt^{-1}meter^{-1}$ is the permittivity
of free space. Then the final form of the Lorentz force experienced by an
electric dipole moment in the presence of electric field is

\begin{equation}
\vec{F}_{elec}=\frac{qd}{4\pi ~\epsilon _{o}r^{3}}\left( 2\cos \Phi ~\hat{r}%
\mathbf{+}\sin \Phi ~\hat{\Phi}\mathbf{+0~}\hat{\Theta}\right) .
\end{equation}%
Similarly as we did for the magnetic force, we can write the radial,
transverse, and normal components $\left( R_{elec},T_{elec},~N_{elec}\right)
$ of the electric force,

\begin{equation}
R_{elec}=-\frac{q}{m}\frac{d}{4\pi ~\epsilon _{o}r^{3}}\left( \omega _{e}-%
\dot{\Theta}\right) \sin ^{2}\Phi .
\end{equation}

\begin{equation}
T_{elec}=\vec{F}_{elec}\cdot \hat{t}~=\frac{q}{m}\frac{d}{4\pi ~\epsilon
_{o}r^{3}}\frac{r^{3}}{\sqrt{\mu ~a\left( 1-e^{2}\right) }}\left( \omega
_{e}-\dot{\Theta}\right) ~\dot{\Phi}\sin \Phi ~\cos \Phi .
\end{equation}

\begin{eqnarray}
N_{elec} &=&\vec{F}_{elec}\cdot \hat{n}~ \\
&=&\frac{q}{m}\frac{d}{4\pi ~\epsilon _{o}r^{3}}\frac{r^{2}}{\sqrt{\mu
~a\left( 1-e^{2}\right) }}\left( \omega _{e}-\dot{\Theta}\right) ~\dot{\Theta%
}~\sin ^{2}\Phi ~\cos \Phi .  \notag
\end{eqnarray}%
Similarly, we can write the components of the Lorentz force experienced by
an electric field as a function of orbital elements as follows.

\begin{equation}
R_{elec}=-\frac{q}{m}\frac{d}{4\pi ~\epsilon _{o}r^{3}}\left[
\begin{array}{c}
\omega _{e}\left( 1-\sin ^{2}i\sin ^{2}\left( \omega ^{\ast }+f\right)
\right) - \\
\sqrt{\mu /a^{3}\left( 1-e^{2}\right) ^{3}}~\cos i~\left( 1+e~\cos f\right)
^{2}%
\end{array}%
\right] .
\end{equation}

\begin{eqnarray}
T_{elec} &=&\frac{q}{m}\frac{d}{4\pi ~\epsilon _{o}r^{2}}\times  \label{35}
\\
&&\left( \omega _{e}-\sqrt{\mu /a^{3}\left( 1-e^{2}\right) ^{3}}\frac{\cos
i~\left( 1+e~\cos f\right) ^{2}}{1-\sin ^{2}i\sin ^{2}\left( \omega ^{\ast
}+f\right) }\right) \sin ^{2}i\cos \left( \omega ^{\ast }+f\right) \sin
\left( \omega ^{\ast }+f\right) .  \notag
\end{eqnarray}

\begin{eqnarray}
N_{elec} &=&\frac{q}{m}\frac{d}{4\pi ~\epsilon _{o}r^{3}}\times  \label{36}
\\
&&\left( \omega _{e}-\sqrt{\mu /a^{3}\left( 1-e^{2}\right) ^{3}}\frac{\cos
i~\left( 1+e~\cos f\right) ^{2}}{1-\sin ^{2}i\sin ^{2}\left( \omega ^{\ast
}+f\right) }\right) \sin i~\cos i\sin \left( \omega ^{\ast }+f\right) ~.
\notag
\end{eqnarray}%
Assuming that the spacecraft is equipped with a charged surface (screen) of
area $S$ with the electric charge $q=\int_{{S}}\sigma \,dS$ distributed over
the surface with density $\sigma $. Therefore, as in Tikhonov et al. (2011),
we can write the torque of these forces relative to the spacecraft's center
of mass as follows.

\begin{equation}
{\vec{T}}_{{L}}=\vec{T}_{mag}+\vec{T}_{elec}=\int_{S}{\sigma }{\vec{\rho}}%
\times (\vec{E}+\vec{V}\times {\vec{B})}{dS},
\end{equation}%
where $\vec{\rho}$\ is the radius vector of the screen's element $dS$
relative to the spacecraft's center of mass and $\vec{V}$ is the velocity of
the element $dS$ relative to the geomagnetic field. Finally, the torque due
to Lorentz force can be written as follows

\begin{equation}
\vec{T}_{mag}=\vec{\rho}_{{0}}\times {A}^{T}\left(
R_{mag},T_{mag},~N_{mag}\right) ^{T},\vec{T}_{elec}=\vec{\rho}_{{0}}\times {A%
}^{T}\left( R_{elec},T_{elec},~N_{elec}\right) ^{T},
\end{equation}

\begin{equation}
{\vec{\rho}}_{{0}}{=x}_{{0}}{i}_{{b}}{+y}_{{0}}{j}_{{b}}{+z}_{{0}}k_{{b}}{=}%
q^{-1}\int_{{S}}\sigma \,\vec{\rho}\,dS
\end{equation}%
$\vec{\rho}_{{0}}$\ is the radius vector of the charged center of a
spacecraft relative to its center of mass and $A^{{T}}$ is the transpose of
the matrix ${A.}$

\subsection{ Geomagnetic field model and its Torque}

In this paper we are using non-tilted dipole for the geomagnetic field. Let
a dipole magnetic field be $\vec{B}=(B_{1},B_{2},B_{3}),$ and the magnetic
moment be $\vec{M}=(m_{1},m_{2},m_{3})$ of the spacecraft. Therefore the
torque due to the geomagnetic field is
\begin{equation}
\vec{T}_{M}=\vec{M}\times \vec{B}
\end{equation}%
As in Wertz (1978) we can write geomagnetic field and the total magnetic
moment of the orbital system directed to the tangent of the orbital plane,
normal to the orbit, and in the direction of the radius respectively as
below:
\begin{eqnarray}
B_{1} &=&\frac{B_{0}}{2r^{3}}\sin \theta _{m}^{\prime }\,[3\cos (2f-\alpha
_{m})+\cos \alpha _{m}],\text{ }B_{2}=-\frac{B_{0}}{2r^{3}}\,\cos \theta
_{m}^{\prime }\,, \\
B_{3} &=&\frac{B_{0}}{2r^{3}}\sin \theta _{m}^{\prime }\,[3\sin (2f-\alpha
_{m})+\sin \alpha _{m}].
\end{eqnarray}

\begin{equation}
m_{1}=m\,\sin \theta _{m}^{\prime }\cos \alpha _{m}\,{\beta _{1}}%
,~m_{2}=m\,\sin \theta _{m}^{\prime }\sin \alpha _{m}\,{\beta _{2}},\text{ }%
m_{3}=m\,\cos \theta _{m}^{\prime }\,{\beta _{3}},
\end{equation}%
where $B_{0}=-8\times 10^{-15}$, $\theta _{m}^{\prime }=168.6^{\circ }$ is
the co-elevation of the dipole, $\alpha _{m}=109.3^{\circ }$, is the east
longitude of the dipole and $f$ is the true anomaly measured from ascending
node and $m$ is the magnitude of the total magnetic moment.

\section{ Equations of the attitude motion}

The nonlinear differential equation called Euler-Poisson \ equations are
used to describe the attitude orientation of the spacecraft.

\begin{equation}
{\ }\dot{\vec{\omega}}I+\vec{\omega}\times \vec{\omega}I=\vec{T}_{G}+\vec{T}%
_{M}+\vec{T}_{L},  \label{54}
\end{equation}%
\begin{equation}
\dot{\vec{\alpha}}+\vec{\alpha}\times \vec{\omega}\mathbf{=-}\Omega \vec{%
\gamma},\dot{\vec{\beta}}+\vec{\beta}\times \vec{\omega}=0,\,\,\,\,\,\,\,%
\dot{\vec{\gamma}}+\vec{\gamma}\times \vec{\omega}=\Omega \vec{\alpha}
\end{equation}%
where $\vec{T}_{G}=3\Omega ^{2}\vec{\gamma}\times \vec{\gamma}I$ is the well
known formula of the gravity gradient torque,

$I={diag}(A,B,C)$ is the inertia matrix of the spacecraft, $\Omega $ is the
orbital angular velocity, $\vec{\omega}$\textbf{\ }is the angular velocity
vector of the spacecraft. According to Wertz (1978) the angular velocity of
the spacecraft in the inertial reference frame is $\vec{\omega}=(p,q,r)%
\mathrm{,}$ where
\begin{equation}
p={\dot{\psi}\sin \theta \sin \phi +\dot{\theta}\,\cos \phi ,}q={\dot{\psi}%
\sin \theta \cos \phi -\dot{\theta}\,\sin \phi ,}r={\dot{\psi}\cos \theta +%
\dot{\phi}.}
\end{equation}

\subsection{Equations of motion in the pitch direction}

In this section the attitude motion of the spacecraft in the pitch direction
is considered, i.e. $\psi =\phi =0,~\theta \neq 0.$ Applying this condition
in equation (\ref{54}), we can derive the second order differential equation
of the motion in pitch direction.

\begin{eqnarray}
{A}\frac{d^{2}\theta }{dt^{2}} &=&(C-B)(3\Omega ^{2}-1){\sin \theta \cos
\theta }+z_{0}{\sin \theta }(~N_{mag}-kT_{mag}) \\
&&+z_{0}{\cos \theta }(kN_{mag}-T_{mag}){+}z_{0}{\sin \theta }%
(~N_{elec}-kT_{elec})  \notag \\
&&+z_{0}{\cos \theta }(kN_{elec}-T_{elec})+{m_{2}B}_{3}{\sin \theta +m_{3}B}%
_{2}{\cos \theta .}  \notag
\end{eqnarray}

Let $y_{0}=k~z_{0},$where $k$ is arbitrary number.%
\begin{eqnarray}
A\ddot{\theta} &=&\left( 3\Omega ^{2}-1\right) (C-B)\sin \theta \cos \theta
+z_{0}\sin \theta (N_{mag}-kT_{mag})  \notag \\
&&+z_{0}\cos \theta (kN_{mag}-T_{mag})+z_{0}\sin \theta (N_{elec}-kT_{elec})
\notag \\
&&+z_{0}\cos \theta (kN_{elec}-T_{elec})+B_{2}m_{3}\sin \theta
+B_{3}m_{2}\cos \theta  \notag \\
&=&g(k,z0,\alpha ^{\ast }=q/m,\theta ),  \label{pitchEq}
\end{eqnarray}%
where $N_{mag},T_{mag},N_{elec},T_{elec}$ are given in equations (\ref{26}),
(\ref{27}), (\ref{35}) and (\ref{36}) respectively.

\begin{figure}[tbp]
\centering
\resizebox{130mm}{!}{\includegraphics {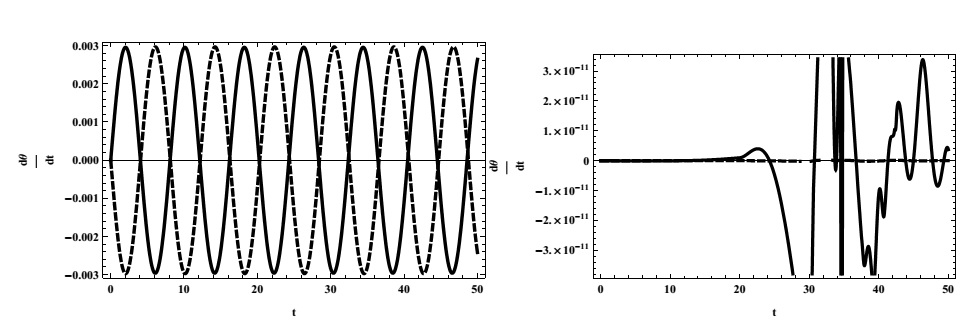}}
\caption{{Oscillation in $\frac{d\protect\theta}{dt}$ due to (\textbf{left})
magnetic part from Lorentz Torque and (\textbf{right}) due to electric part
from Lorentz Torque. The dotted line corresponds to $\protect\alpha^* =-0.1$
and the continuous line corresponds to $\protect\alpha^*=0.1$}}
\label{Oscillationthetadot1}
\end{figure}

\begin{figure}[t]
\centering
\resizebox{130mm}{!}{\includegraphics {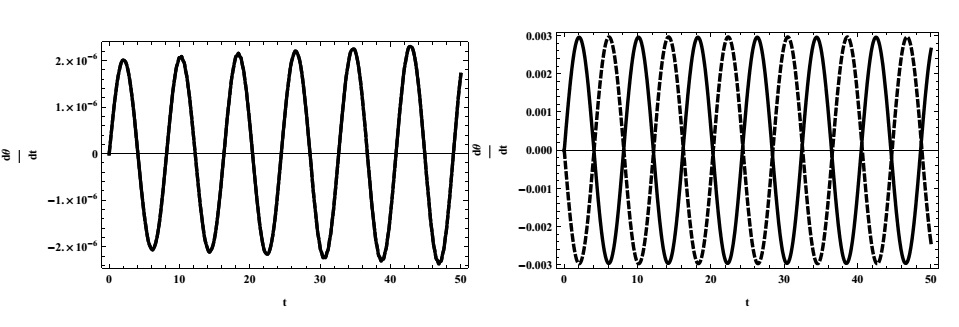}}
\caption{{Oscillation in $\frac{d\protect\theta }{dt}$ due to (\textbf{left}%
) geomagnetic torque  and (\textbf{right}) Torque due to total Lorentz force
and geomagnetic field. The dotted line corresponds to $\protect\alpha ^{\ast
}=-0.1$ and the continuous line corresponds to $\protect\alpha ^{\ast }=0.1$%
. Both the lines for $\protect\alpha ^{\ast }=\pm 0.1$ are overlapping.}}
\label{Oscillationthetadot3}
\end{figure}

A comparison in the oscillation of {\ }$\overset{.}{{\theta }}$ is given in
figures (\ref{Oscillationthetadot1}) to (\ref{Oscillationthetadot3}). It is
obvious from the first two figures that the most significant amount of
torque is coming from the magnetic part of the Lorentz torque which is of
the order $10^{-3}$. The effect from electric part of the Lorentz torque is
of the order $10^{-11}$ which is very small. The contribution from
geomagnetic torque is of the order $10^{-6}$. \ The {oscillation in }$%
\overset{.}{{\theta }}${\ due to total Torque is of the order }$10^{-3},$ as
shown in figure (\ref{Oscillationthetadot3} right). As the contribution from
the electric part of Lorentz force and geomagnetic field is very small
compared to the magnetic part of Lorentz force therefore it doesn't show up
in figure (\ref{Oscillationthetadot3} left). These figures are drawn for
fixed values of $B=0.7$, $C=0.1$, $\alpha ^{\ast }=\pm 1$.

\subsection{Derivation of equilibrium solutions in the pitch direction and
their linear stability analysis}

\begin{figure}[tbp]
\centering%
\resizebox{130mm}{!}{\includegraphics
{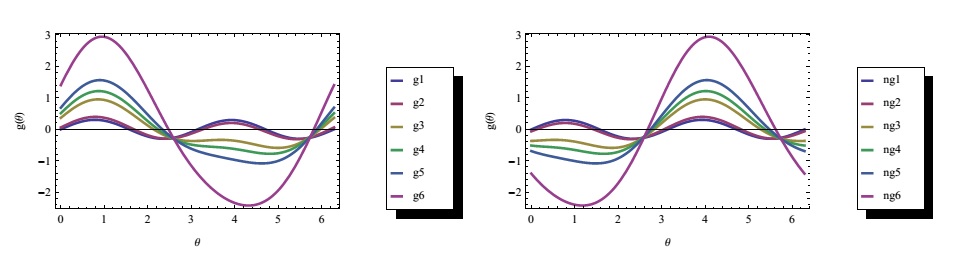}}
\caption{Progression of equilibrium solutions when (\textbf{left}) g1) $(%
\protect\alpha ^{\ast },k)=(0.1,1)$, g2) $(\protect\alpha ^{\ast },k)=(2,2)$%
, g3) $(\protect\alpha ^{\ast },k)=(5,7)$, g4) $(\protect\alpha ^{\ast
},k)=(7,7)$, g5) $(\protect\alpha ^{\ast },k)=(7,10)$, g6) $(\protect\alpha %
^{\ast },k)=(10,15)$ and (\textbf{right}) when ng1) $(\protect\alpha ^{\ast
},k)=(-0.1,1)$, ng2) $(\protect\alpha ^{\ast },k)=(-2,2)$, ng3) $(\protect%
\alpha ^{\ast },k)=(-5,7)$, ng4) $(\protect\alpha ^{\ast },k)=(-7,7)$, ng5) $%
(\protect\alpha ^{\ast },k)=(-7,10)$, ng6) $(\protect\alpha ^{\ast
},k)=(-10,15)$ }
\label{rootprogressionTHETA-1}
\end{figure}

\begin{figure}[tbp]
\centering\resizebox{70mm}{!}{\includegraphics {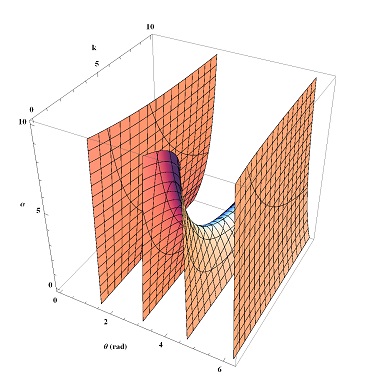}}
\caption{Implicit plot of $g(k,\protect\alpha ^{\ast },\protect\theta )=0$
when $z_{0}=1$.}
\label{allrootstheta}
\end{figure}

In this section the existence and stability of equilibrium position in the
pitch direction of a general shape spacecraft under the influence of
gravitational torque, Lorentz torque, and geomagnetic torque will be
discussed. The stability of the equilibrium solutions derived will be
discussed both analytically and numerically. To find the equilibrium
solutions, take the right hand side of equation (\ref{pitchEq}) equal to
zero which reduces to the following equation for $B=0.7,C=0.1,$ $%
a=6900km,i=51^{\circ },e=0.001,$ and $f=60^{\circ }.$

\begin{eqnarray}
g(k,z0,\alpha ^{\ast },\theta ) &=&(2.34\times
10^{-7}+(0.015+0.008k)z_{0}\alpha ^{\ast })\cos \theta  \label{pitcheq2} \\
&&+(1.56\times 10^{-6}+(0.008+0.015k)z_{0}\alpha ^{\ast })\sin \theta
+0.3\sin 2\theta =0.  \notag
\end{eqnarray}%
It is not possible to solve equation (\ref{pitcheq2}) in closed form as $%
\theta =f(k,z_{0},\alpha ^{\ast })$ therefore numerical techniques are used
to identify all the roots of equation (\ref{pitcheq2}). As equation (\ref%
{pitchEq}) is derived by taking$\ y=kz_{0},$ therefore without loss of
generality we take $z_{0}=1.$ For $0<\alpha ^{\ast }<1$ and $0<k<1,$ we have
five equilibrium solutions at $\theta \approx \frac{n\pi }{2},n=0,1,2,3,4$
when $\theta \in \lbrack 0,2\pi ].$ As $g(k,\alpha ^{\ast },\theta )$ is a
periodic function of period $2\pi $ therefore it is sufficient to
investigate the equilibrium solutions from $0$ to $2\pi .$ For $0<\alpha
^{\ast }<1$ and $k\leq 100$ there are five equilibrium solutions which
reduces to three or two when $k>100$. For sufficiently high values of $%
\alpha ^{\ast },$ the number of equilibrium solutions can be reduced to
three for even smaller values of $\alpha ^{\ast }.$ To see the progression
of roots from five to three see figures (\ref{rootprogressionTHETA-1}) where
$g(k,\alpha ^{\ast },\theta )$ is plotted for various fixed values of $k$
and $\alpha ^{\ast }.$ To completely describe the progression of the number
of equilibrium positions in $[0,2\pi ]$ from five to two a 3D implicit plot
of $g(k,\alpha ^{\ast },\theta )=0$ is given in figure (\ref{allrootstheta}%
). It can easily be seen that for high enough values of $\alpha ^{\ast }$
and $k$, the number of equilibrium points reduces to two. It is also obvious
from these figures that the equilibrium positions does not always remain at $%
\theta \approx \frac{n\pi }{2},n=0,1,2...$. By the comparison of figure (\ref%
{rootprogressionTHETA-1} left) and figure (\ref{rootprogressionTHETA-1}
right) it is evident that the equilibrium positions are not the same for
positively and negatively charged spacecrafts. It remains to be seen if this
or the other parameters such as $\alpha ^{\ast }$ or $k$ effect the
stability of the equilibrium points.

To discus the linear stability of the equilibrium points identified above we
use the standard procedure of linearization and convert equation (\ref%
{pitchEq}) to a system of two first order equations. We then find the
eigenvalues of the jacobian matrix from the equation given below.

\begin{equation}
\lambda ^{2}-\frac{\partial g(k,z0,\alpha ^{\ast },\theta )}{\partial \theta
}=0,  \label{Charactristic}
\end{equation}%
where%
\begin{eqnarray}
\frac{\partial g(k,z0,\alpha ^{\ast },\theta )}{\partial \theta } &=&A^{-1}({%
\cos \theta (1.56\times 10^{-6}+10^{-3}(8+}{15k)z_{0}\alpha ^{\ast })}
\label{G-theta} \\
&&{+(B-C)\cos (2\theta )}{-\sin \theta (2.34\times 10^{-7}+10^{-3}(15}{%
+8k)z_{0}\alpha ^{\ast })}.  \notag
\end{eqnarray}%
It is clear from equation (\ref{Charactristic}) that there are only two
types of eigenvalues possible. If $g_{\theta }(k,z0,\alpha ^{\ast },\theta
)>0$ there will be two eigenvalues one of which is negative and one
positive. A positive eigenvalue always imply instability. If $g_{\theta
}(k,z0,\alpha ^{\ast },\theta )<0$ the eigenvalues obtained will be
imaginary with a zero real part which means the equilibrium point in
question will be spectrally stable. Initially we will investigate the
equilibrium points obtained above for $A=1,B=0.7,C=0.1,z_{0}=1.$

\begin{eqnarray}
\frac{\partial g(k,\alpha ^{\ast },\theta )}{\partial \theta } &=&g_{\theta
}={\cos \theta (1.56\times 10^{-6}+10^{-3}(8+15k)\alpha ^{\ast })}{+0.6\cos
(2\theta )}  \notag \\
&&-{\sin \theta (2.34\times 10^{-7})}{+10^{-3}\sin \theta (15+8k)\alpha
^{\ast }.}
\end{eqnarray}

\begin{figure}[tbp]
\centering%
\resizebox{130mm}{!}{
        \includegraphics{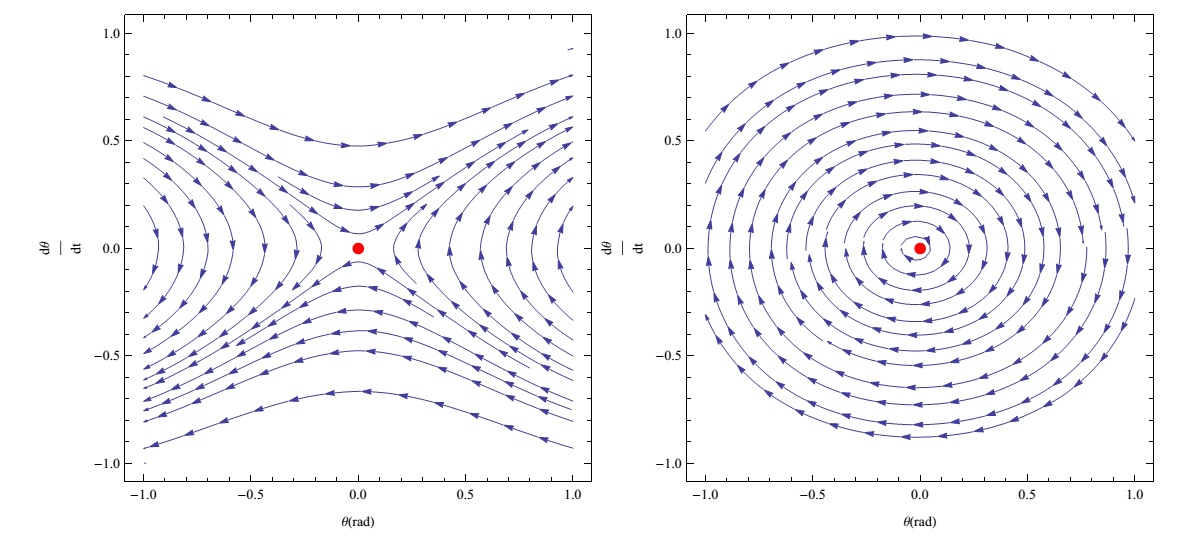}}
\caption{Trajectory in the $\protect\theta $-$\frac{d\protect\theta }{dt}$
phase plane when \textbf{(left) }$\protect\alpha ^{\ast
}=0.01,k=1,z_{0}=1,A=1,B=0.7,C=0.1$,\textbf{(right) }$\protect\alpha ^{\ast
}=-1,k=100,z_{0}=1,A=1,B=0.7,C=0.1$. }
\label{theta1}
\end{figure}
The values of $g_{\theta }|_{\theta =0}$ and $g_{\theta }|_{\theta =2\pi }$
remain positive for all positive values of $k$ and $\alpha ^{\ast }$ which
implies that the equilibrium position at $\theta =0$ and $\theta =2\pi $ are
unstable. The value of $g_{\theta }|_{\theta =\pi /2}$ is negative for all
positive values of $k$ and $\alpha ^{\ast }$ which implies that the
equilibrium position at $\theta =\pi /2$ will be stable. By similar
argument, the equilibrium position at $\theta =\pi $ will be stable if $%
\alpha ^{\ast }$ satisfy the following inequality.%
\begin{equation*}
\alpha ^{\ast }>\frac{0.6}{0.008+0.015k}=\alpha _{1}^{\ast }.
\end{equation*}%
This also means that $\theta =\pi $ will always be unstable if the
spacecraft is negatively charged as the right hand side of the above
inequality is always positive. To check the stability of the remaining four
equilibrium positions when $\alpha ^{\ast }<0,$ Let $\alpha ^{\ast }=-\alpha
_{p}$ such that $\alpha _{p}>0.$ It can easily be shown that the equilibrium
position at $\theta =0$ and $\theta =2\pi $ will be stable if $\alpha
_{p}<\alpha _{1}^{\ast }.$ For example when $k=1,$ $\alpha _{p}$ must be
smaller than $26.09.$ Similarly, for the equilibrium position at $\theta
=\pi /2$ to be stable for negatively charged spacecraft $\alpha _{p}$ must
satisfy the following inequality.%
\begin{equation*}
\alpha _{p}<\frac{0.6}{0.008k+0.015}.
\end{equation*}%
It can be safely concluded from this discussion that the sign and amount of
charge on the spacecraft plays a significant role in the stability of the
equilibrium positions. A typical trajectory in the $\theta $-$\frac{d\theta
}{dt}$ phase plane around $\theta =0$ is given in figure ( \ref{theta1}) for
$\alpha ^{\ast }=0.01,-1,$ $k=1,100,$ $z_{0}=1,$ $A=1,$ $B=0.7,$ $C=0.1$. It
can be seen that all the trajectories are moving away from $\theta =0$ when $%
\alpha ^{\ast }=0.01,$ which indicate instability. In the second case it is
stable.

\begin{figure}[tbp]
\centering
\resizebox{130mm}{!}{
        \includegraphics{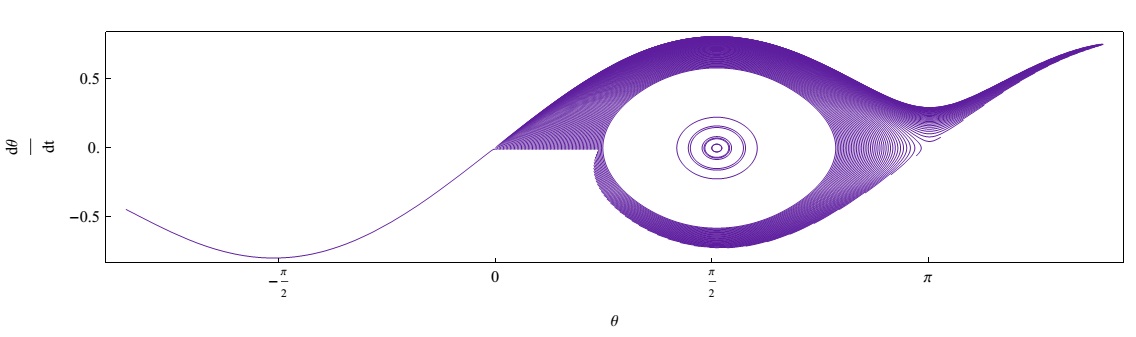}}
\caption{A set of orbits with initial positions close to $(\protect\theta ,%
\dot{\protect\theta )}\rightarrow (0+,0+)$ in the $\protect\theta $-$\frac{d%
\protect\theta }{dt}$ phase plane when $\protect\alpha ^{\ast
}=0.01,k=1,z_{0}=1,A=1,B=0.7,C=0.1$ }
\label{theta2}
\end{figure}
\begin{figure}[tbp]
\centering
\resizebox{130mm}{!}{
        \includegraphics{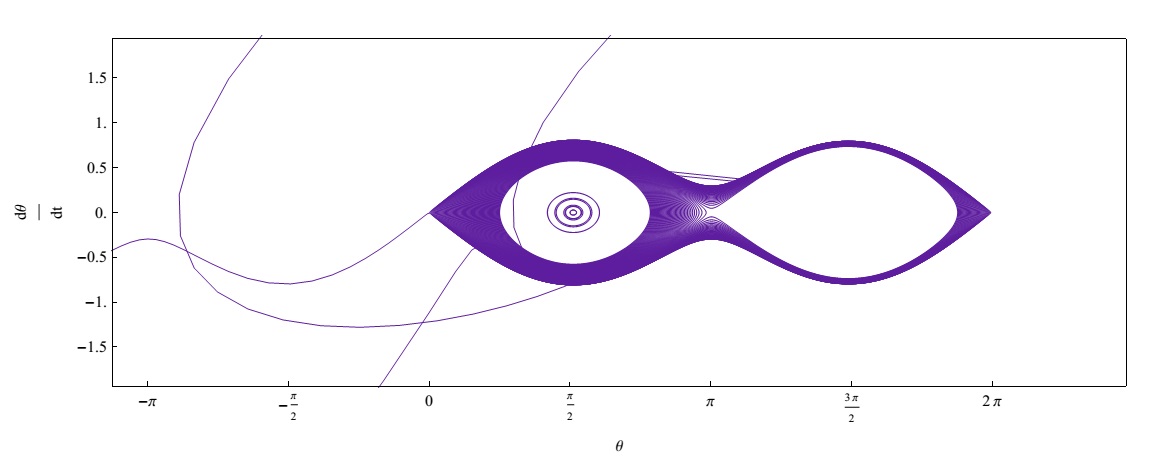}  }
\caption{Same orbits as in figure (\protect\ref{theta2}) but integrated for
much longer time.}
\label{theta3}
\end{figure}

\begin{figure}[tbp]
\centering
\resizebox{130mm}{!}{
        \includegraphics{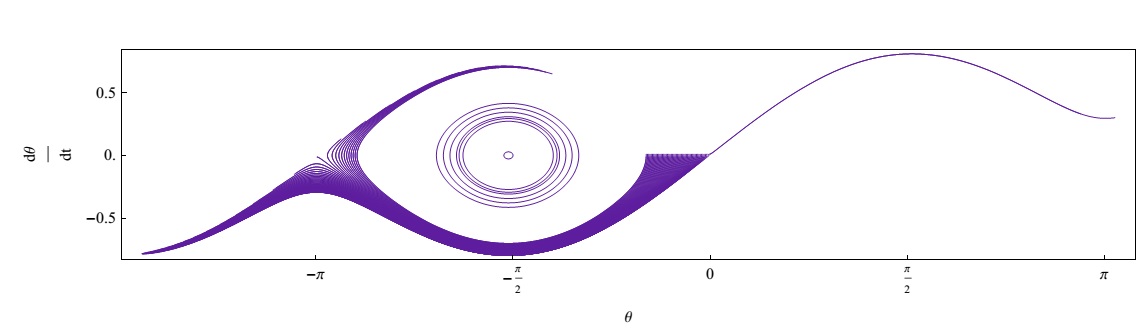} }
\caption{A set of orbits with initial positions close to $(\protect\theta ,%
\dot{\protect\theta})\rightarrow (0-,0-)$ in the $\protect\theta $-$\frac{d%
\protect\theta }{dt}$ phase plane when $\protect\alpha ^{\ast
}=0.01,k=1,z_{0}=1,A=1,B=0.7,C=0.1$ }
\label{theta4}
\end{figure}
\begin{figure}[tbp]
\centering
\resizebox{140mm}{!}{
        \includegraphics{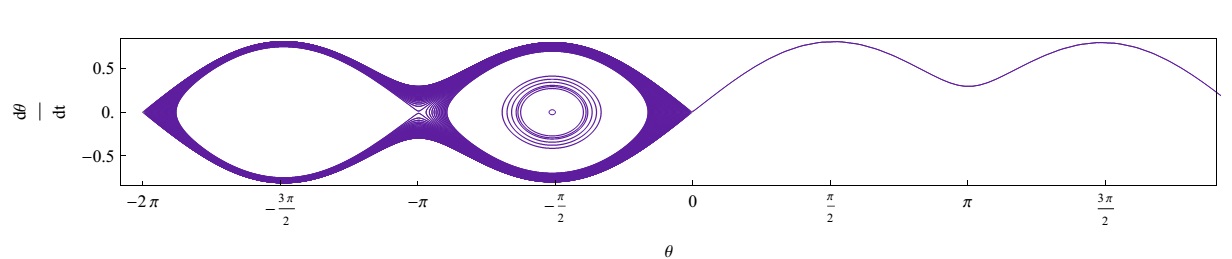}}
\caption{Same orbits as in figure (\protect\ref{theta4}) but integrated for
much longer time.}
\label{theta5}
\end{figure}

To understand the long term behavior of orbits around the equilibrium
positions a set of orbits with initial positions close to $(\theta ,\dot{%
\theta )}\rightarrow (0+,0+)$ and $(\theta ,\dot{\theta )}\rightarrow (\frac{%
\pi }{2},0)$ are given in figures ( \ref{theta2}, \ref{theta3}) in the $%
\theta $-$\frac{d\theta }{dt}$ phase plane when $\alpha ^{\ast
}=0.01,k=1,z_{0}=1,A=1,B=0.7,C=0.1$. These orbits are allowed to evolve for
a short period of time and their trajectories are traced in figure ( \ref%
{theta2}). It can be seen that the orbits starting close to $0+$ (close to 0
and positive) are immediately captured by the nearby stable equilibrium at $%
\frac{\pi }{2}.$ The orbits which start near $(\theta ,\dot{\theta )}%
\rightarrow (\frac{\pi }{2},0)$ remain in elliptic orbit around $(\frac{\pi
}{2},0).$ When these orbits are allowed to evolve for a longer period of
time some of the orbits near $(0+,0+)$ are being captured by the nearby
stable equilibrium at $\frac{3\pi }{2}$ and a couple of orbits escapes. The
orbits close to $(\frac{\pi }{2},0)$ remain in near circular orbit about the
center which is a strong evidence of the existence of periodic orbits around
$(\frac{\pi }{2},0).$ Similar analysis is performed for orbits with initial
positions close to $(\theta ,\dot{\theta )}\rightarrow (0-,0-)$ and $(\theta
,\dot{\theta )}\rightarrow (-\frac{\pi }{2},0).$ It can be seen in figures ( %
\ref{theta4}, \ref{theta5}) that the orbits starting close to $0-$ (close to
0 and negative) are immediately captured by the nearby center at $-\frac{\pi
}{2}.$ Some of them when integrated for a much longer period of time gets
captured by the center at $\frac{-3\pi }{2}.$ The orbits which start around $%
(-\frac{\pi }{2},0)$ remain in near circular orbits around $(-\frac{\pi }{2}%
,0).$ Similar behavior is observed around all the spectrally stable
equilibriums. Therefore we can safely conjecture that around each stable
equilibrium position there is a family of periodic orbits.

As mentioned earlier and shown in figures (\ref{rootprogressionTHETA-1}) and
(\ref{allrootstheta}) the number of equilibrium points \ when $0<\theta
<2\pi $ reduce from five to three and in some cases two for higher values of
$\alpha ^{\ast }$ and $k.$ For example when $B=0.7,C=0.1,k=10,z_{0}=1,$ and $%
\alpha ^{\ast }=7$ there are two equilibrium points at $\theta =2.32$
(stable)$,5.89$ (unstable)$.$ If $\alpha ^{\ast }=-7$ i.e. the spacecraft is
negatively charged, the position of the two equilibriums are changed and the
stability reversed. When $|B-C|<1$, the positions of the equilibrium points
are almost identical to what we have shown above. Its effect on stability is
explained below.

\begin{enumerate}
\item $\theta =0,2\pi $: When $B\geq C$ and $\alpha ^{\ast }>0,$ the
equilibrium positions at $\theta =0,2\pi $ will be stable. But when $\alpha
^{\ast }<0$ these two equilibrium positions are unstable.

\item $\theta =\frac{\pi }{2}$: For $B\geq C$ the equilibrium position at $%
\theta =\frac{\pi }{2}$ is always stable$.$ However, when $B<C$ the value of
$\alpha ^{\ast }$ have to be significantly high in which case $\theta =\frac{%
\pi }{2}$ will no more be an equilibrium position.

\item $\theta =\pi :$When $B\leq C$ and $\alpha ^{\ast }>0,$ $\theta =\pi $
is stable. For $\ $\ a negatively charged spacecraft $B<C$ is a necessary
condition for the stability of the equilibrium position at $\theta =\pi .$
Therefore when $B>C,$ the value of $\alpha ^{\ast }$ have to be
significantly high in which case $\theta =\pi $ will no more be an
equilibrium position.

\item $\theta =\frac{3\pi }{2}:B>C$ is a necessary and sufficient condition
for the stability of the equilibrium position at $\theta =\frac{3\pi }{2}$
unless $\alpha ^{\ast }$ is very large and negative in which case $\theta =%
\frac{3\pi }{2}$ will no more be at equilibrium position.
\end{enumerate}

\begin{figure*}[tbp]
\centering%
\resizebox{130mm}{!}{
        \includegraphics{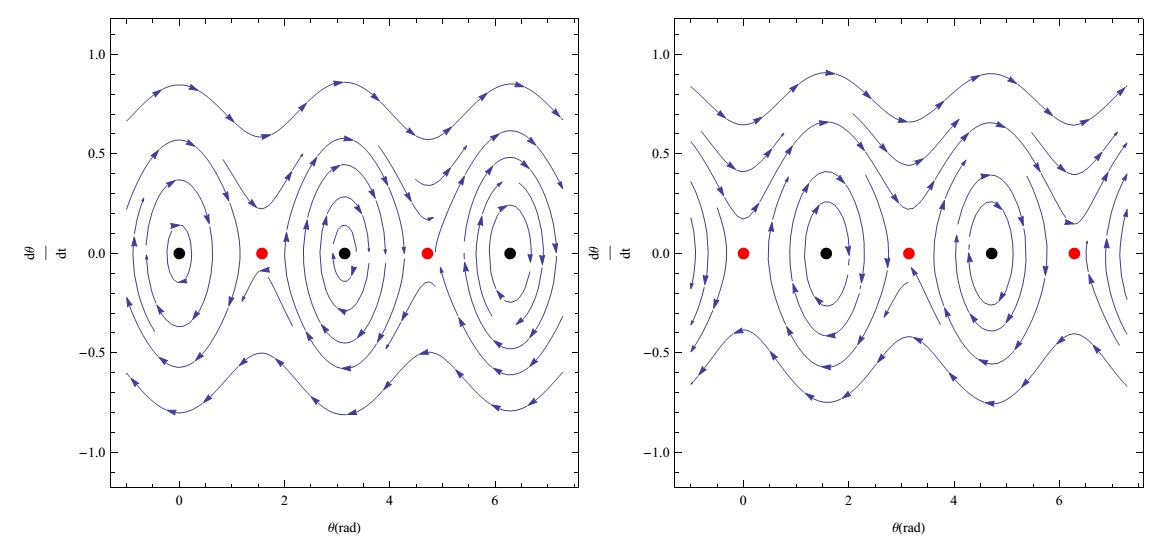}}
\caption{Trajectory in the $\protect\theta $-$\frac{d\protect\theta }{dt}$
phase plane when $\protect\alpha =0.01,k=1,z_{0}=1$ and \textbf{\ (Left)}: $%
B=0.5,C=0.9$. \textbf{(Right)}: $B=0.9,C=0.5$. Black dots correspond to
stable equilibrium points and red dots correspond to unstable equilibrium
points.}
\label{theta6}
\end{figure*}

In summary, when $0<\alpha ^{\ast }<1$ and $B<C,$ the equilibrium positions
at $\theta =0,\pi ,2\pi $ are stable and at $\theta =\frac{\pi }{2},\frac{%
3\pi }{2}$ are unstable and when $B>C$ the nature of the five equilibrium
positions is reversed. To demonstrate this behavior a typical example is
given in figure (\ref{theta6})

\subsection{Equations of motion in the roll direction}

In this section we study the attitude motion of the spacecraft in the roll
direction, i.e. $\psi =\theta =0,~\phi \neq 0.$ Applying this condition in
Euler equation of the attitude motion of the spacecraft, we obtain the
second order differential equation of the motion in roll direction.%
\begin{eqnarray*}
{C}\frac{d^{2}{\phi }}{dt^{2}} &=&\Omega ^{2}(A-B){\sin \phi \cos \phi }%
+x_{0}{\sin \phi }(kT_{mag}-~R_{mag})+x_{0}{\cos \phi }(T_{mag}-kR_{mag}) \\
&&+x_{0}{\sin \phi }(kT_{elec}-~R_{elec})+x_{0}{\cos \phi (T_{elec}}{%
-~kR_{elec})+m_{1}B}_{2}{\cos \phi -{m_{2}B}_{1}\sin \phi .}
\end{eqnarray*}

Let ${y}_{{0}}=k{x}_{{0}}~$meter then,

\begin{eqnarray}
{\ C}\frac{d^{2}{\phi }}{dt^{2}} &=&\Omega ^{2}(A-B){\sin \phi \cos \phi }%
+x_{0}{\sin \phi }(kT_{mag}-~R_{mag})+x_{0}{\cos \phi }(T_{mag}-kR_{mag})
\notag \\
&&{+}x_{0}{\sin \phi }(kT_{elec}-~R_{elec})+x_{0}{\cos \phi (T_{elec}}{%
-~kR_{elec})\cos \phi +m_{1}B}_{2}{\cos \phi -{m_{2}B}_{1}\sin \phi }  \notag
\\
&=&h(\alpha ^{\ast },k,{\phi ,x}_{0},{A,B).}  \label{rollEq}
\end{eqnarray}

\subsection{Derivation of equilibrium solutions in Roll direction and their
linear stability analysis}

In this section the existence and stability of equilibrium positions in the
roll direction of a general shape spacecraft under the influence of
gravitational torque, Lorentz torque, and geomagnetic torque will be
discussed. The stability of the equilibrium positions derived will be
discussed both analytically and numerically. To find the equilibrium
positions, take the right hand side of equation (\ref{rollEq}) equal to zero
which reduces to the following equation for $A=0.1,B=0.7,C=1,$ $x_{0}=1,$ $%
a=6900km,i=51^{\circ },e=0.001,$ and $f=60^{\circ }.$
\begin{eqnarray}
h_{1}(k,\alpha ^{\ast },\phi )=(1.036\times 10^{-7}+(-0.012+9.41k)\alpha
^{\ast }) &&\times \cos \phi +(-1.13\times 10^{-7}  \notag \\
+9.41\times 10^{-10}\alpha ^{\ast }-0.012k\alpha ^{\ast })\sin \phi
&&-3.63\times 10^{-7}\sin (2\phi )=0.
\end{eqnarray}

It is not possible to solve equation $h_{1}(k,\alpha ^{\ast },\phi )=0$ in
closed form as $\phi =f(k,\alpha ^{\ast })$ therefore numerical techniques
are used to identify all the roots of equation $h_{1}(k,\alpha ^{\ast },\phi
)=0$ which are the desired equilibrium solutions. Let $k=1.$ For $\alpha
^{\ast }\in (-2.15\times 10^{-5},2.15\times 10^{-5})$ i.e for a very small
amount of charge, there are four equilibrium solutions and for higher values
of $\alpha ^{\ast }$ there are two equilibrium solutions.

\begin{enumerate}
\item $\phi _{1}\in (0,0.4)$ when $\alpha ^{\ast }\in (-2.15\times
10^{-5},2.15\times 10^{-5}).$

\item $\phi _{2}\in (1.47,1.89)$ when $\alpha ^{\ast }\in (-2.15\times
10^{-5},2.15\times 10^{-5}).$

\item $\phi _{3}\in (2.80,3.18)$ when $\alpha ^{\ast }\in (-2.15\times
10^{-5},2.15\times 10^{-5}).$

\item $\phi _{4}\in (4.37,4.73)$ when $\alpha ^{\ast }\in (-2.15\times
10^{-5},2.15\times 10^{-5})$

\item $\phi _{5}=2.36$ when $\left\vert \alpha ^{\ast }\right\vert
>2.15\times 10^{-5}$

\item $\phi _{6}=5.5$ when $\left\vert \alpha ^{\ast }\right\vert
>2.15\times 10^{-5}$
\end{enumerate}

In the above example $B>A.$ Now we switch the values of $A$ and $B$ to have $%
B<A$ and find the location of the equilibrium positions. In this case we
still have four equilibrium solutions when $\alpha ^{\ast }\in (-2.15\times
10^{-5},2.15\times 10^{-5})$ and two when $\left\vert \alpha ^{\ast
}\right\vert >2.15\times 10^{-5}.$ All the equilibrium positions when $%
A=0.7, $ $B=0.1,k=1,C=1,x_{0}=1,$ $a=6900km,i=51^{\circ },e=0.001,$ and $%
f=60^{\circ }$ are listed below.

\begin{enumerate}
\item $\phi _{7}\in (1.23,1.58)$ when $\alpha ^{\ast }\in (-2.15\times
10^{-5},2.15\times 10^{-5}).$

\item $\phi _{8}\in (3.12,3.47)$ when $\alpha ^{\ast }\in (-2.15\times
10^{-5},2.15\times 10^{-5}).$

\item $\phi _{9}\in (4.69,5.04)$ when $\alpha ^{\ast }\in (-2.15\times
10^{-5},2.15\times 10^{-5}).$

\item $\phi _{10}\in (5.97,6.32)$ when $\alpha ^{\ast }\in (-2.15\times
10^{-5},2.15\times 10^{-5})$

\item $\phi _{11}=2.36$ when $\left\vert \alpha ^{\ast }\right\vert
>2.15\times 10^{-5}$

\item $\phi _{12}=5.5$ when $\left\vert \alpha ^{\ast }\right\vert
>2.15\times 10^{-5}$
\end{enumerate}

\begin{figure}[tbp]
\centering
\resizebox{65mm}{!}{
        \includegraphics{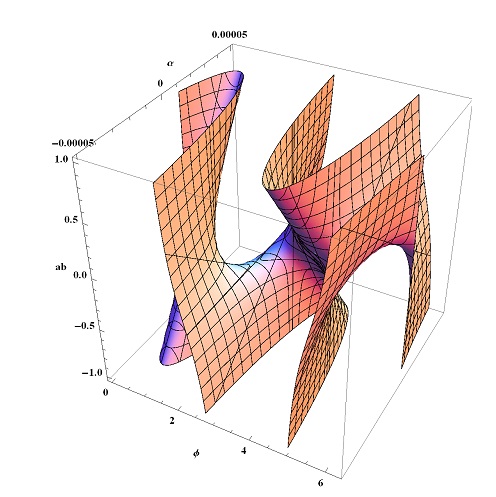} }
\caption{Implicit plots of ( $h(\protect\alpha ,\protect\phi ,ab)=0$. }
\label{roll1}
\end{figure}

\begin{figure}[tbp]
\centering
\resizebox{130mm}{!}{
        \includegraphics{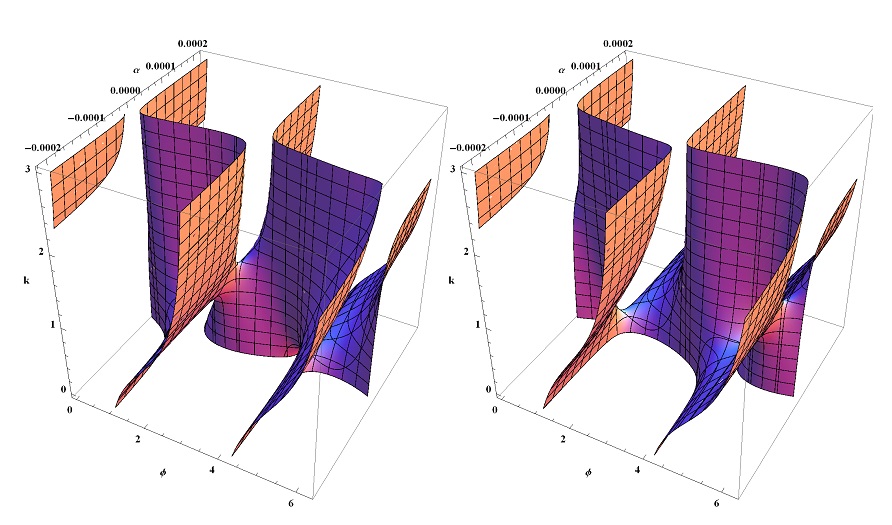}}
\caption{Implicit plots of $h(\protect\alpha ,\protect\phi ,k,ab)=0$ when (%
\textbf{left}) $ab=-0.3$ and (\textbf{right}) $ab=0.3$.}
\label{roll2b}
\end{figure}

As ${y}_{0}=kx_{0}$ we can take $x_{0}=1.$ To reduce the dimensions, without
loss of generality, we define $ab=A-B$ and rewrite $h_{1}(k,\alpha ^{\ast
},\phi ,ab)$ as below.%
\begin{eqnarray*}
h_{1}(k,\alpha ^{\ast },\phi ,ab) &=&\left( -1.13\times
10^{-7}+10^{-3}(6k-15)\alpha ^{\ast }\right) {\cos }\phi +1.03\times 10^{-7}
\\
&&+10^{-3}(6-15k)\alpha ^{\ast }){\sin }\phi +6.05\times 10^{-7}ab{\sin }%
(2\phi ).
\end{eqnarray*}%
It can be seen from figure (\ref{roll1}) that there are four equilibrium
solutions for small values of $\alpha ^{\ast }$ and all values of $ab$ when $%
k=1$. For higher values of $\alpha ^{\ast }$ there are only two equilibrium
solutions at $\phi =2.3$ and $\phi =5.5$ for all values of $ab$. We have
seen above that for $k=1$, the changing values of $ab$ and $\alpha ^{\ast }$
have significant effect on the existence of equilibrium solutions in the
roll direction. To see this for $k\neq 1$, we plot $h_{1}(k,\alpha ^{\ast
},\phi ,ab)=0$ for fixed values of $ab=0.3,$ and $ab=-0.3$ in figure (\ref%
{roll2b}). It is clear from figure (\ref{roll2b}) that with the changing
value of $k$ the position of equilibrium changes significantly but the
numbers of equilibrium positions remains four as before both for negative
and positive values of $ab$ when $\alpha ^{\ast }$ is small. The effect of $%
ab$ is significant when $k<1.$ For higher values $\alpha ^{\ast }$ the
number of equilibrium positions remains to be two but their positions change
with the changing values of $ab$ and $k,$ see figures (\ref{roll1} and \ref%
{roll2b}).

\begin{figure*}[tbp]
\centering
\resizebox{130mm}{!}{
        \includegraphics{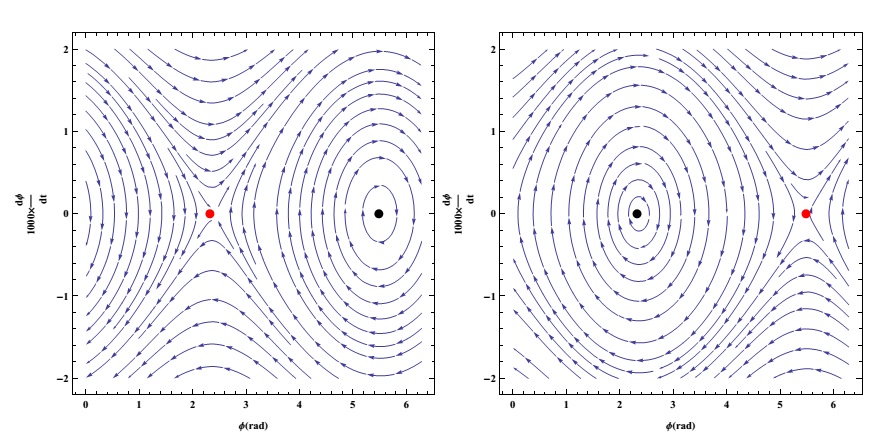}        }
\caption{Trajectory in the $\protect\phi $-$\frac{d\protect\phi }{dt}$ phase
plane when \textbf{(left) }$\protect\alpha ^{\ast }=0.1,k=1,z_{0}=1,ab=-0.6$,%
\textbf{(right) }$\protect\alpha ^{\ast }=-0.1,k=1,z_{0}=1,ab=-0.6$. }
\label{phiphase11}
\end{figure*}

To study the stability of the equilibrium position derived we use the same
method which was used for the pitch direction. We write equation (\ref%
{rollEq}) as a system of two first order equations, linearize them, and find
the eigenvalues of the jacobian matrix from the equation given below.

\begin{equation}
\lambda ^{2}-h_{\phi }(\alpha ^{\ast },k,\phi ,ab)=0.  \label{CharEqphi}
\end{equation}%
It can be seen from equation (\ref{CharEqphi}) that there are only two types
of eigenvalues possible. If $h_{\phi }(\alpha ^{\ast },k,\phi ,ab)>0$ there
exist two eigenvalues one of which is negative and one is positive.
Therefore $h_{\phi }(\alpha ^{\ast },k,\phi ,ab)>0$ becomes a sufficient
condition for instability. If $h_{\phi }(\alpha ^{\ast },k,\phi ,ab)<0$ the
equilibrium point in question will be spectrally stable or a stable center.
We will investigate the equilibrium points obtained above for $%
ab=0.6,ab=-0.6,k=1,x_{0}=1$,$~a=6900km,i=51^{\circ },e=0.001,f=60^{\circ },$
and write $h_{\phi }(\alpha ^{\ast },\phi ,ab)$ as below.%
\begin{eqnarray*}
h_{\phi }(\alpha ^{\ast },\phi )|_{ab=-0.6} &=&(-1.13\times
10^{-7}-0.01\alpha ^{\ast })\cos \phi -7.26\times 10^{-7}(\cos \phi )^{2} \\
&&+(-1.04\times 10^{-7}+0.01\alpha ^{\ast }+7.26\times 10^{-7}\sin \phi
)\sin \phi .
\end{eqnarray*}%
The equilibrium positions at $\phi _{1}$ and $\phi _{3}$ are stable as in
these cases $h_{\phi }(\alpha ^{\ast },\phi )|_{ab=-0.6}<0$. Similarly $\phi
_{2}$ and $\phi _{4}$ are unstable as in these cases $h_{\phi }(\alpha
^{\ast },\phi )|_{ab=-0.6}>0.$ By similar arguments $\phi _{5}$ will be an
unstable equilibrium if the spacecraft is positively charged and $\phi _{6}$
will be unstable if the spacecraft is negatively charged. Similarly, when $%
ab=0.6,$ $\phi _{7}$ and $\phi _{9}$ are stable, $\phi _{8}$ and $\phi _{10}$
are unstable, $\phi _{11}$ is stable when $\alpha ^{\ast }<-1.11\times
10^{-6}$ and $\phi _{12}$ is stable when $\alpha ^{\ast }>3.25\times
10^{-7}. $ A typical example is given in figure (\ref{phiphase11}) when $%
ab=\pm 0.6$. The equilibrium at $\phi =2.36$ is stable when $\alpha ^{\ast
}=-0.1$ and unstable when $\alpha ^{\ast }=0.1$. Similarly, the equilibrium
at $\phi =5.5 $ is unstable when $\alpha ^{\ast }=-0.1$ and stable when $%
\alpha ^{\ast }=0.1$.
\begin{figure*}[tbp]
\centering
\resizebox{130mm}{!}{
        \includegraphics{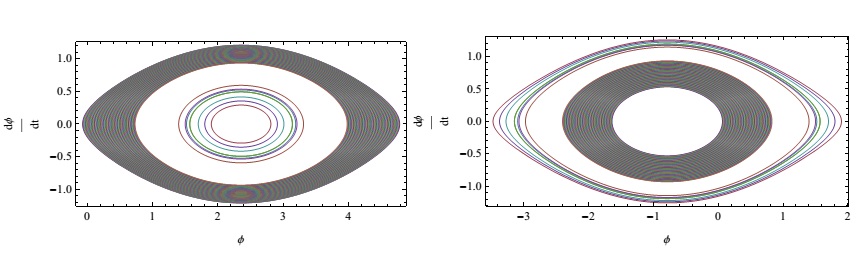}        }
\caption{A set of orbits with initial positions close to the equilibrium
positions when $e=0.1,ab=0.6$ and (left) $\protect\alpha ^{\ast }=-0.1$,
(right) $\protect\alpha ^{\ast }=0.1$ }
\label{phiorbit1}
\end{figure*}

\begin{figure*}[tbp]
\centering
\resizebox{130mm}{!}{
        \includegraphics{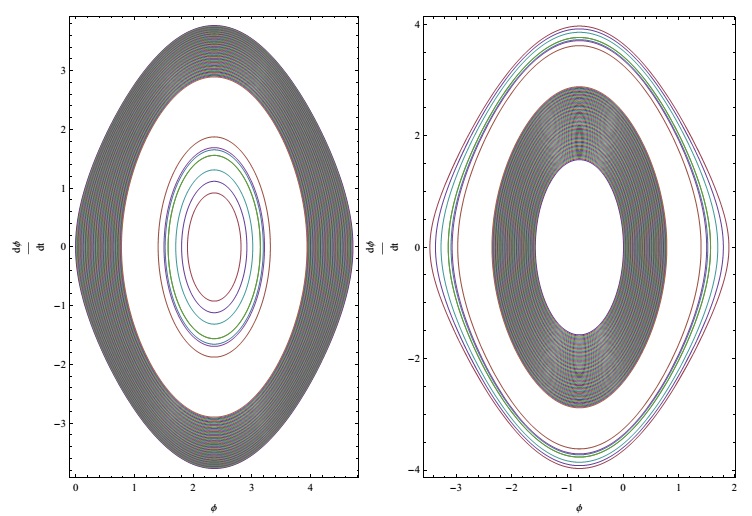}        }
\caption{A set of orbits with initial positions close to the equilibrium
positions when $e=0.1,ab=0.6$ and (left) $\protect\alpha ^{\ast }=-1$,
(right) $\protect\alpha ^{\ast }=1$ }
\label{phiorbit2}
\end{figure*}

To understand the long term behavior of orbits around the equilibrium
positions, a set of orbits with initial positions close to the equilibrium
are given in figures ( \ref{phiorbit1}, \ref{phiorbit2}) in the $\phi $-$%
\frac{d\phi }{dt}$ phase plane when $\alpha ^{\ast }=\pm 0.1,\pm 1,$ and $%
k=1,z_{0}=1,ab=0.6,e=0.1$. These orbits are allowed to evolve for a long
period of time and their trajectories are traced in figures (\ref{phiorbit1}%
, \ref{phiorbit2}). The orbits in figure (\ref{phiorbit1}left) are given for
$\alpha ^{\ast }=-0.1$ and it can be seen that all the orbits are captured
by the equilibrium position at $\phi =2.36$ which is a stable equilibrium.
The orbits which are closer to the stable equilibrium position remain in
perfect periodic orbit while the orbits which are not so close have an
elliptic orbit in the vicinity of the equilibrium position but are not
necessarily periodic. For $\alpha ^{\ast }=0.1$ in figure ( \ref{phiorbit1}%
right), the equilibrium position at $\phi =2.36$ is unstable. Hence the same
orbits are captured by another nearby stable equilibrium at $\phi =-0.723$
which is a mirror image of the stable equilibrium at $\phi =5.56$. When the
same orbits are integrated for $\alpha ^{\ast }=\pm 1$, similar behavior is
observed. Also, similar behavior is observed around all the stable
equilibriums. Therefore we can safely conjecture that around each stable
equilibrium position there is a family of periodic orbits.

\section{Conclusions}

The paper discussed the attitude stabilization of a charged spacecraft
moving in an elliptic orbit using Lorentz torque. The Lorentz torque is
developed in two parts $T_{mag}$ and $T_{elec}.$ $T_{mag}$ is the Lorentz
torque which is experienced by magnetic field and $T_{elec}$ is the Lorentz
Torque experienced by an electric dipole moment in the presence of electric
field. The model we developed incorporates all Lorentz torques as a function
of orbital elements and the radius vector of the charged center of the
spacecraft relative to it's center of mass. We investigated, both
analytically and numerically, the existence and stability of equilibrium
positions both in pitch and roll directions. In the pitch direction there
are a total of five equilibrium points at $\theta =n\pi /2,n=0,1,2,3,4$ when
$-1<\alpha ^{\ast }=q/m<1$, $0<k<1$ and $\theta \in \lbrack 0,2\pi ]$ .
Their stability is analyzed for changing values of the charge to mass ratio,
$\alpha ^{\ast }$, and it is shown that $\alpha ^{\ast }$ effect the
stability and existence of equilibrium positions. The equilibrium positions
at $\theta =0,2\pi $ are unstable for $\alpha ^{\ast }>0$ when $B=0.7$ and $%
C=0.1.$ These two equilibrium positions are stable when $B=0.1$ and $C=0.7.$
These equilibrium positions are also stable for $\alpha ^{\ast }<0.$ We have
shown that the sign and amount of charge play a significant role in
determining the equilibrium positions and their stability. In the case of
roll direction we have four equilibrium points when $\alpha ^{\ast }\in
(-2.15\times 10^{-5},2.15\times 10^{-5})$ and only two equilibrium positions
when $\alpha ^{\ast }\notin (-2.15\times 10^{-5},2.15\times 10^{-5}).$ It is
demonstrated both analytically and numerically that almost all the
equilibrium positions depend on the values and sign of charge to mass ratio
both in terms of existence and stability. In the same way as in pitch
direction, the equilibrium positions which are stable for $A<B$ becomes
unstable when $A>B$ and vice versa. This is not true in general but this
happens in most of the cases. Here $A,B$ and $C,$ refers to the components
of moment of inertia of the spacecraft.

\end{document}